\documentclass{aa}  

\usepackage{graphicx}
\usepackage{txfonts}
\usepackage{caption}

\begin{document} 

\newcommand{\pw}{PW20 }

   \title{NuSTAR view of heavily absorbed AGN: The $R-N_\text{H}$ correlation}

   \author{C. Panagiotou
          \and
          R. Walter
          \and
          S. Paltani
          }

   \institute{
              Astronomy Department, University of Geneva, Chemin d’Ecogia 16, 1290 Versoix, Switzerland\\
              \email{Christos.Panagiotou@unige.ch}
         }

   \date{Received ***; accepted ***}

 
  \abstract
   {The nature of the putative torus and the outer geometry of active galactic nuclei (AGN) are still rather unknown and the subject of active research. Improving our understanding of them is crucial for developing a physical picture for the structure of AGN.}
   {The main goal of this work is to investigate the outer geometry of AGN by studying the observed hard X-ray spectrum of obscured sources. We primarily aim at researching the reflected emission in these sources.}
   {To that end, we analysed archived \textit{NuSTAR} observations of a sample of nearby AGN, whose X-ray emission has been found to be heavily absorbed, with $10^{23}<N_\text{H}<2.5\cdot 10^{23}\text{ cm}^{-2}$; the upper limit on $N_\text{H}$ was necessary due to the analysis we followed and the data quality. Fitting their emission with both a phenomelogical and a physical model, we investigated the relation between reflection and absorption.}
   {The strength of reflected emission, as well as the equivalent width of the Fe K$\alpha$ line, correlates with the absorption column density, which can be explained with a clumpy torus origin for the reflection in these sources. The shape of the observed correlation is found to be well reproduced when the effects of a clumpy torus with a variable filling factor are simulated. A similar increase in reflection seems to be featured even by sources with larger absorption, reaching the Compton thick ($N_\text{H}>1.5\cdot 10^{24}\text{ cm}^{-2}$) regime. }
  {}


\maketitle



\section{Introduction}

It is now well established that active galactic nuclei (AGN) are powered by the accretion of matter onto a supermassive black hole located in their centres. Gravitational energy is liberated as matter spirals around the potential well in the form of an accretion disc, which emits quasi-thermal radiation, mainly in the UV and optical wavebands \citep{1973A&A....24..337S}. The X-ray source, often referred to as the corona, is thought to be significantly smaller than the disc and to be located in the immediate vicinity of the black hole, although its exact geometry and physics are still unclear \citep[e.g.][]{2013ApJ...769L...7R}.

The unified model of AGN \citep[e.g.][]{1993ARA&A..31..473A}, which was initially proposed to explain the differences in the optical line spectrum between Seyfert 1 and 2 galaxies, suggests that a dusty doughnut-shape region, the so-called torus, surrounds the central engine. As a result, if our line of sight intersects this region, we observe an obscured AGN; otherwise, the inner engine is directly viewed. In this way, the variety of observed AGN is attributed to a single parameter, the system's inclination angle. Initial support for the unified model was provided by the detection of broad Balmer lines in the polarisation spectrum of Seyfert 2 galaxies \citep{1985ApJ...297..621A} and by the broad, although not universal, agreement between UV and X-ray absorption in AGN \citep{1991PASJ...43..195A, 2012MNRAS.426.1750M}.

However appealing the simple unification might be, it was realised early on that such a homogeneous structure would not be stable \citep{1988ApJ...329..702K}. In addition, observations of source eclipses in the X-rays \citep[e.g.][]{2002ApJ...571..234R, 2003MNRAS.342L..41L, 2014MNRAS.439.1403M} strongly suggest that the absorption is not due to a homogeneous circumnuclear structure. These considerations, among others, led to the idea of a clumpy torus. 

Further evidence of the torus clumpiness has been obtained by infrared studies \citep[see, for example,][for a review]{2015ARA&A..53..365N}. For instance, a continuous torus would result in characteristic differences between the infrared spectra of AGN of different classes, which are not regularly observed. Instead, a clumpy distribution of clouds could explain the similary of mid-infrared spectra between different AGN \citep{2008ApJ...685..160N}. \cite{2008ApJ...685..147N, 2008ApJ...685..160N} developed a model that simulates the effects of a clumpy torus in the infrared spectrum of AGN. This model has been found to reproduce the observed spectra well and to be able to provide constraints on the torus's properties \citep[e.g.][]{2013ApJ...764..159L}.

An independent approach to constraining the outer geometry of AGN consists in the investigation of its emission spectrum in hard X-rays. In this energy range, the Compton hump, or reflection hump, which is produced  by the interaction of the primary continuum emission with the surrounding matter \citep[e.g.][]{1991MNRAS.249..352G}, contributes significantly to the observed spectrum \citep[e.g.][]{2016MNRAS.458.2454L}. Therefore, the study of the reflection hump may provide information about the nature and geometry of the matter surrounding the central engine. To that end, one may also investigate the Fe K$\alpha$ emission line, a prominent feature of AGN spectra \citep[e.g.][]{2007MNRAS.382..194N}, which is produced when X-rays illuminate the nearby matter as well. In recent years, a few studies have been successful in reproducing the X-ray spectral properties of AGN using a clumpy torus model \citep{2019A&A...629A..16B, 2020ApJ...897....2T}.

Furthermore, previous studies have found that the strength of the Compton hump emission with respect to the primary continuum, $R$, is significantly large in highly obscured AGN, which hints at the existence of a clumpy torus. For example, \cite{2011A&A...532A.102R} were the first to report a high fraction of reflected emission in obscured sources with an absorption column density of $N_\text{H} > 10^{23} \text{cm}^{-2}$ when studying the average International Gamma-Ray Astrophysics Laboratory \citep[INTEGRAL;][]{2003A&A...411L...1W} spectra of local AGN. This feature was explained as an increase in the covering factor of the torus or as a result of the torus clumpiness. 

The strong reflection in mildly obscured objects was further verified by \cite{2013ApJ...770L..37V} and later by \cite{2016A&A...590A..49E} using stacked Burst Alert Telescope \citep[BAT;][]{2005SSRv..120..143B} spectra of nearby AGN. Moreover, \cite{2017ApJ...849...57D} found tentative evidence for a positive $R$-$N_\text{H}$ correlation when considering the average Nuclear Spectroscopic Telescope Array \citep[\textit{NuSTAR};][]{2013ApJ...770..103H} spectra of AGN residing at redshift $z=0.04-3.21$. These AGN were observed as part of the \textit{NuSTAR} extragalactic survey. However, their results were not supported by the analysis of \cite{2018ApJ...854...33Z}, who studied the brightest sources of the \textit{NuSTAR} extragalactic survey programme and found evidence for an anti-correlation between $R$ and $N_\text{H}$, although at a low significance level.

\cite{2019A&A...626A..40P} found a positive correlation between reflection and absorption using individual \textit{NuSTAR} spectra of nearby AGN. More recently, \cite{2020ApJ...901..161K} found a similar trend when analysing the broadband X-ray spectra of 19 nearby Seyfert 2 galaxies. The detected correlation was shown to be well reproduced by a simple power law with a slope of 0.87. Furthermore, \cite{2020MNRAS.tmp.2543B} studied the \textit{Chandra} spectra of AGN detected in four \textit{Chandra} deep fields and found a weak increase in reflection with absorption. However, limited by the data quality, the validity of this increase could not be confirmed.

In \cite{2020A&A...640A..31P} (\pw \hspace*{-4pt}, hereafter), we investigated the X-ray reflected emission in a large sample of nearby AGN. We showed that sources at different obscuration states feature a similar range of reflection strengths, which seems in tension with the assumption that these sources differ solely in their inclination angle. In unobscured sources (with $N_\text{H} < 5 \times 10^{22} \text{cm}^{-2}$), the reflection strength was concluded to correlate with the photon index, which was suggested to be driven by the motion of the X-ray source. No such correlation was found, though, in the case of more obscured sources, hinting at a different behaviour of the reflected emission in these objects. Here, we study in more detail the reflected emission of these obscured sources.

Motivated by our previous results, in this work we examine the existence of an $R$-$N_\text{H}$ correlation in a large sample of Seyfert galaxies, constraining our analysis to sources with $N_\text{H} > 10^{23} \text{cm}^{-2}$. Such a correlation is important since it could provide constraints on the geometry and nature of the torus in these objects. We evaluate statistically the reality of the correlation between reflection and absorption using different spectral models and we investigate whether the effects of a clumpy torus can reproduce the apparent correlation. The sample and the reduction procedure are described in Sect. \ref{sect:data_red}. Section \ref{sect:results} presents the results of the spectral modelling, which are further discussed in Sect. \ref{sect:discuss}. We summarise our main findings in Sect. \ref{sect:conclud}.


\section{Data sample and reduction}
\label{sect:data_red}

\begin{figure}
  \centering
  \includegraphics[width=0.65\linewidth,height=0.55\linewidth, trim={10 40 10 70}, clip]{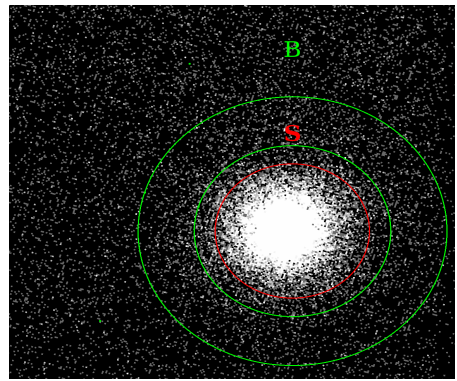}
  \caption[10]{Full-band \textit{NuSTAR} image of the sky region towards SWIFT J0048.8+3155. The red circle, denoted by `S', and the green annulus, denoted by `B', indicate the source and background regions respectively, which were used in the data analysis. The source region has a radius of 110 arcseconds. The image corresponds to the \textit{NuSTAR} observation 60160026002 and its colour distribution is in logarithmic scale. }
  \label{fig:j0048d8_img}
\end{figure}

We used the sample presented in PW20. It consists of all the sources classified as Seyfert 1 or 2 in the 70-month \textit{Swift}/BAT catalogue \footnote{https://swift.gsfc.nasa.gov/results/bs70mon/ \newline Seyfert 1 and 2 galaxies are categorised as Class 4 and 5, respectively, in this catalogue.} \citep{2013ApJS..207...19B} that had public archival \textit{NuSTAR} data by April 2019. They span a redshift range from 0.002 to 0.118. A detailed description of the sample characteristics and the data reduction is given in \pw \hspace*{-4pt}. In brief, all \textit{NuSTAR} observations were reduced using the \textit{NuSTAR} Data Analysis Software (NuSTARDAS). Spectral energy distributions were produced from the cleaned event files for all sources, while part of the observations was not considered due to the passage of \textit{NuSTAR} through the South Atlantic Anomaly area. The source spectra were extracted from a circular region with the sources placed in their centres, while the background spectra were extracted mostly from an annular region encircling the source regions. The radii of the two regions were chosen with the aim to maximise the signal-to-noise ratio and to avoid source contamination in the background emission. An example of these regions for the Seyfert 2 SWIFT J0048.8+3155 is shown in Fig. \ref{fig:j0048d8_img}. All the source spectra were then rebinned to reach 25 source counts in each energy bin to allow the use of Gaussian statistics \citep[e.g.][]{1971smep.book.....E}.

\pw divided the sources in three classes based on their measured absoption column densities: unobscured ($N_\text{H} < 5 \times 10^{22} \text{cm}^{-2}$), lightly obscured ($5 \times 10^{22} \text{cm}^{-2} < N_\text{H} < 10^{23} \text{cm}^{-2}$), and mildly obscured sources (MOB, hereafter) with $N_\text{H} > 10^{23} \text{cm}^{-2}$. There were also several sources with apparent reflection dominated spectra, which would naturally correspond to even larger values of $N_\text{H}$, that were excluded from the previous analysis. Due to its shape, fitting the \textit{NuSTAR} spectrum of these latter sources did not allow for a reliable determination of the corresponding spectral parameters, and, therefore, these objects could not be used in investigating the different correlations.

Here, we consider only the MOB sources\footnote{In Sect. \ref{sect:rvsnh}, we further constrain our sample to sources with $N_\text{H} < 25 \times 10^{22} \text{cm}^{-2}$ to avoid the introduction of bias due to the exclusion of the reflection dominated sources.}. There were 40 such sources in total \footnote{There were 13 additional MOB sources, for which the photon index had to be fixed during the fit. We do not consider these sources in the current work to avoid introducing biases.}, of which three Seyfert 1 and 37 Seyfert 2 galaxies. Although considering only MOB sources constrains the considered range of $N_\text{H}$, it was necessary within the scope of the conducted analysis. The reflected emission of sources with lower absorption seemed to have a different behaviour, while, as already mentioned, sources of larger $N_\text{H}$ could not be considered due to their spectral shapes and the model's degeneracy. A discussion of how our results extrapolate outside the considered $N_\text{H}$ range and on how this work could be extended to larger absorption is given in Sect. \ref{sect:discuss}.


\section{Results}

\label{sect:results}

\subsection{Spectral fitting}

We fitted the spectra using two different models in order to examine if our results are model-independent. As described in more detail below, we firstly used a simple phenomelogical model to reproduce the observed spectra, while a more physical model was then employed. For each of the considered models, the model fitting was conducted using the \textit{XSPEC} software \citep{1996ASPC..101...17A} and $\chi^2$ statistics, while the abundances of \cite{2009ARA&A..47..481A} were assumed when the phenomelogical model was used. The spectra of both \textit{NuSTAR} detectors, FPMA and FPMB, were fitted simultaneously with a cross-normalisation between them left free to be minimised during the fit. Unless otherwise mentioned, all the errors stated in this work correspond to a 1-$\sigma$ confidence interval. In estimating the model luminosity of the sources, a $\Lambda$CDM cosmology was assumed with $H_0 = 70 \text{ km}/ \text{s}/ \text{Mpc}$, $\Omega_\text{tot} = 1$,  and $\Omega_\Lambda = 0.73$ .

\subsubsection{Using a phenomelogical model}
\label{sect:pexrav_res}

All the spectra were initially fitted with a simple phenomelogical model. This model has already been used in \pw \hspace*{-4pt} and is described again here for consistency. 

The \textit{pexrav} model \citep{1995MNRAS.273..837M} was used to simulate the intrinsic AGN emission as a power law with a cutoff at high energies and its reflection from a slab medium with infinite optical depth. The modification of AGN emission due to photo-electric absorption in the host galaxy was modelled by the \textit{zphabs} model. The Galactic absorption was not considered because of its small value; $N_\text{H}^\text{Gal} \lesssim 10^{22} \text{cm}^{-2}$. Finally, a Gaussian line was added to the model in order to account for the Fe K$\alpha$ emission line, centred at 6.4 keV. 

The normalisation of the Gaussian line was let free during the fit, while its width was kept fixed at 0.05 keV, since it could not be resolved with \textit{NuSTAR} resolution. The validity of this was checked a posteriory; letting the line width free did not improve the fit significantly for any of the considered sources, except for the Seyfert 2 \textit{SWIFT J1838.4-6524}. This source was found to feature two narrow lines in its spectrum. We fitted this source's spectrum using two narrow Gaussian lines, one fixed at 6.4 keV and one for which the central energy was let free to be minimised. The best-fit energy of the second line was $E = 6.86 \pm 0.06 \text{ keV}$. This line most likely corresponds to emission from highly ionised Fe atoms.

The inclination angle (as measured by the disc symmetry axis) was assumed to be the same for all sources and equal to cos $i$=0.45 in order to reduce the number of degeneracies in the model. Solar abundancies were assumed for the reflecting region. Eventually, in addition to the model normalisations and the cross-normalisation between the two detectors, there were four parameters left free during the fit, the photon index, $\Gamma$, the absorption column density, $N_\text{H}$, the reflection strength, $R$, which quantifies how `strong' the reflected emission is with respect to the continuum power law, and the energy of the cutoff, $E_c$.

\begin{figure}
  \centering
  \includegraphics[width=\linewidth,height=1.5\linewidth, trim={10 40 10 70}, clip]{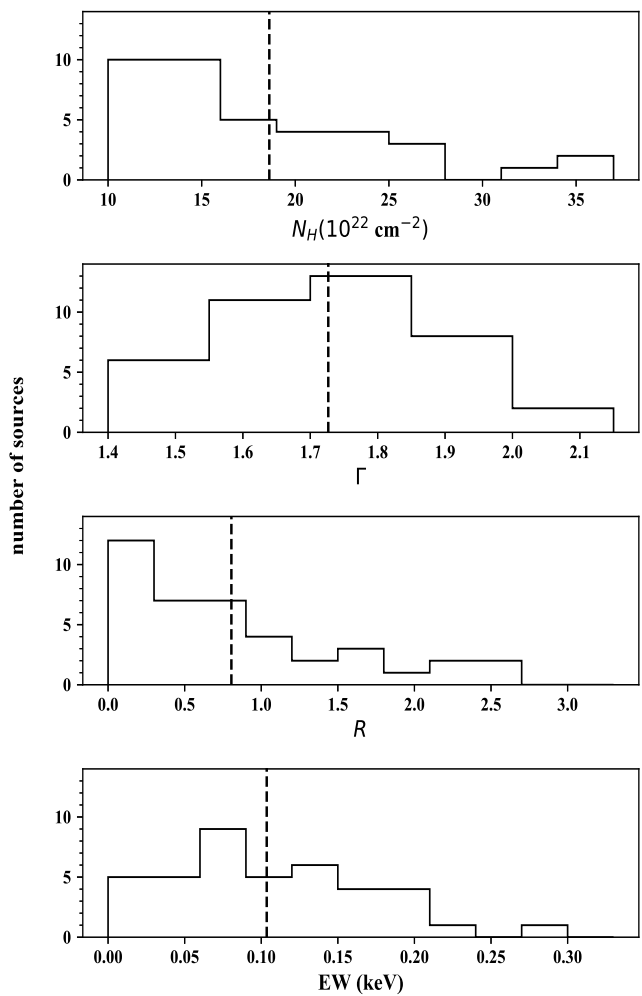}
  \caption[10]{Distribution of the best-fit $N_H$, $\Gamma$, $R$, and Fe K$\alpha$ EW. The vertical dashed line in each plot denotes the average value of the corresponding distribution.}
  \label{fig:bestfit_hist}
\end{figure}

The best-fit results are listed in Table \ref{tab:best_fit_pexrav}. The last column lists the best-fit $\chi^2$ statistic and the corresponding degrees of freedom (dof). All the sources are well fitted with an average reduced $\chi^2$ statistic of <$\chi^2$>$ = 0.97$ and a null hypothesis probability of $P_\mathrm{null}>5\%$ for all the fits. The source distributions of $N_\text{H}$, $\Gamma$, and $R$ are plotted in Fig. \ref{fig:bestfit_hist}. The bottom panel of this figure shows the distribution of the Fe K$\alpha$ equivalent width (EW). The dashed vertical lines in the panels of Fig. \ref{fig:bestfit_hist} denote the average value of each distribution. These are <$N_\text{H}$>$=18.6 \cdot 10^{22}\text{ cm}^{-2}$, <$\Gamma$>$=1.73$, <$R$>$=0.81$, and <$EW$>$=0.10\text{ keV}$.

\begin{table*}
	\centering
	\caption{Best-fit parameters of the \textit{pexrav} model.}
	\label{tab:best_fit_pexrav}
	\begin{tabular}{lllllrr} 
		\hline
		{Source Name}         &  $N_\text{H} $             &  $\Gamma$                & $R$                      &Fe K$\alpha$ EW      &$\chi^2$/df    &BAT class    \\
		                      &  $(10^{22}$cm$^{-2})$      &                          &                          &   (eV)              &               &             \\
		\hline
		SWIFT J0048.8+3155    & $10.9 \pm 0.7$             & $1.65 \pm 0.05$          & $0.27^{+0.11}_{-0.10}$   & $58 \pm 12$         & $838.9/926$   & 5           \\
    SWIFT J0152.8-0329    & $18.6^{+3.0}_{-4.2}$       & $1.78^{+0.13}_{-0.31}$   & $2.22^{+1.68}_{-1.04}$   & $298^{+71}_{-66}$   & $82.2/97$     & 5            \\
    SWIFT J0241.3-0816    & $14.3 \pm 1.2$             & $1.47^{+0.09}_{-0.08}$   & $0.07^{+0.15}_{-0.07}$   & $156 \pm 19$        & $548.9/627$   & 5              \\
    SWIFT J0505.8-2351    & $15.1 \pm 1.4$             & $1.77 \pm 0.10$          & $0.55^{+0.27}_{-0.22}$   & $53^{+19}_{-23}$    & $534.3/549$   & 5              \\
    SWIFT J0526.2-2118    & $19.2^{+2.0}_{-2.1}$       & $1.63^{+0.10}_{-0.13}$   & $0.07^{+0.24}_{-0.07}$   & $35^{+34}_{-28}$    & $295.9/284$   & 5              \\
    SWIFT J0623.9-6058    & $27.4^{+2.7}_{-3.2}$       & $2.06^{+0.19}_{-0.26}$   & $2.45^{+1.87}_{-1.11}$   & $<60$               & $149.4/178$   & 5              \\
    SWIFT J0641.3+3257    & $16.0^{+2.1}_{-2.2}$       & $1.74 \pm 0.13$          & $0.39^{+0.42}_{-0.32}$   & $64^{+45}_{-37}$    & $171.2/192$   & 5              \\
    SWIFT J0742.3+8024    & $13.6 \pm 3.9$             & $1.48^{+0.22}_{-0.23}$   & $<0.32$                  & $<34$               & $113.2/115$   & 4              \\
    SWIFT J0804.2+0507    & $24.1^{+2.5}_{-2.4}$       & $1.65 \pm 0.17$          & $1.33^{+0.69}_{-0.44}$   & $121^{+28}_{-36}$   & $360.6/320$   & 5              \\
    SWIFT J0804.6+1045    & $14.6^{+2.2}_{-1.2}$       & $1.83^{+0.13}_{-0.09}$   & $0.02^{+0.32}_{-0.02}$   & $61^{+38}_{-45}$    & $176.3/167$   & 4              \\
    SWIFT J0843.5+3551    & $35.5^{+5.0}_{-5.6}$       & $1.81^{+0.24}_{-0.28}$   & $0.82^{+1.29}_{-0.72}$   & $<44$               & $63.2/67$     & 5            \\
    SWIFT J0855.6+6425    & $13.7^{+2.5}_{-1.6}$       & $1.94^{+0.18}_{-0.20}$   & $1.75^{+1.74}_{-1.03}$   & $76^{+81}_{-74}$    & $56.5/59$     & 5            \\
    SWIFT J0902.0+6007    & $18.2^{+5.2}_{-4.5}$       & $1.63^{+0.26}_{-0.15}$   & $0.08^{+0.61}_{-0.08}$   & $175^{+96}_{-88}$   & $60.0/45$     & 5            \\
    SWIFT J0926.1+6931    & $14.5 \pm 1.9$             & $1.86 \pm 0.14$          & $0.67^{+0.60}_{-0.41}$   & $111^{+33}_{-31}$   & $246.4/252$   & 4              \\
    SWIFT J1044.1+7024    & $17.8^{+2.3}_{-3.5}$       & $1.76^{+0.14}_{-0.20}$   & $0.67^{+0.80}_{-0.53}$   & $128^{+73}_{-61}$   & $91.4/86$     & 5            \\
    SWIFT J1046.8+2556    & $10.5^{+4.9}_{-4.7}$       & $1.41^{+0.35}_{-0.34}$   & $0.47^{+0.97}_{-0.47}$   & $186^{+85}_{-88}$   & $69.7/67$     & 5            \\
    SWIFT J1049.4+2258    & $39.6^{+4.6}_{-4.4}$       & $1.44^{+0.15}_{-0.18}$   & $<0.23$                  & $126^{+47}_{-49}$   & $183.3/193$   & 5              \\
    SWIFT J1105.7+5854A   & $18.1^{+5.4}_{-7.3}$       & $1.88^{+0.25}_{-0.51}$   & $1.90^{+4.22}_{-1.50}$   & $129^{+99}_{-87}$   & $40.9/46$     & 5            \\
    SWIFT J1217.2-2611    & $12.5^{+1.5}_{-2.6}$       & $1.86^{+0.12}_{-0.19}$   & $0.93 \pm 0.62$          & $<15$               & $158.4/175$   & 5              \\
    SWIFT J1219.4+4720    & $11.4 \pm 1.1$             & $1.68 \pm 0.09$          & $0.11^{+0.21}_{-0.11}$   & $67^{+22}_{-14}$    & $560.8/567$   & 5              \\
    SWIFT J1325.4-4301    & $11.4 \pm 0.1$             & $1.73 \pm 0.01$          & $<0.003$                 & $57 \pm 2$          & $2785.0/2689$ & 5              \\
    SWIFT J1341.5+6742    & $15.9^{+1.6}_{-1.9}$       & $2.03^{+0.03}_{-0.16}$   & $1.70^{+1.30}_{-0.82}$   & $155^{+45}_{-41}$   & $158.9/161$   & 5              \\
    SWIFT J1354.5+1326    & $31.2^{+7.0}_{-7.2}$       & $1.41^{+0.32}_{-0.30}$   & $0.01^{+0.69}_{-0.01}$   & $<63$               & $51.9/70$     & 5            \\
    SWIFT J1457.8-4308    & $14.7^{+3.2}_{-3.9}$       & $1.84^{+0.17}_{-0.32}$   & $2.52^{+1.85}_{-1.24}$   & $89^{+52}_{-54}$    & $83.8/96$     & 5            \\
    SWIFT J1515.0+4205    & $10.9 \pm 1.0$             & $1.89^{+0.05}_{-0.07}$   & $0.89^{+0.41}_{-0.37}$   & $113^{+34}_{-33}$   & $282.7/248$   & 5              \\
    SWIFT J1621.2+8104    & $18.3^{+5.6}_{-5.4}$       & $1.67^{+0.43}_{-0.40}$   & $1.16^{+2.67}_{-1.06}$   & $211^{+66}_{-89}$   & $66.6/71$     & 5            \\
    SWIFT J1717.1-6249    & $20.6^{+0.5}_{-0.4}$       & $1.90 \pm 0.03$          & $1.42^{+0.16}_{-0.15}$   & $78 \pm 9$          & $1107.0/1100$ & 5              \\
    SWIFT J1824.2+1845    & $13.2^{+2.4}_{-3.9}$       & $1.67^{+0.13}_{-0.26}$   & $0.91^{+0.84}_{-0.57}$   & $143^{+55}_{-52}$   & $95.4/118$    & 5             \\
    SWIFT J1824.3-5624    & $22.9^{+3.2}_{-3.1}$       & $1.65 \pm 0.24$          & $2.17^{+1.33}_{-0.85}$   & $201^{+44}_{-42}$   & $231.5/220$   & 5              \\
    SWIFT J1826.8+3254    & $10.1 \pm 1.4$             & $1.77 \pm 0.12$          & $0.48^{+0.32}_{-0.26}$   & $111^{+26}_{-20}$   & $337.3/389$   & 5              \\
    SWIFT J1830.8+0928    & $20.2^{+2.7}_{-2.1}$       & $1.77^{+0.14}_{-0.26}$   & $0.17^{+0.68}_{-0.17}$   & $155^{+77}_{-64}$   & $68.1/73$     & 5            \\
    SWIFT J1838.4-6524    & $22.9 \pm 0.7$             & $1.86 \pm 0.05$          & $1.57^{+0.19}_{-0.17}$   & $112 \pm 10$        & $1230.3/1178$ & 5              \\
    SWIFT J1913.3-5010    & $24.5^{+3.8}_{-3.7}$       & $1.62 \pm 0.22$          & $0.63^{+0.52}_{-0.38}$   & $66^{+43}_{-35}$    & $152.7/189$   & 5              \\
    SWIFT J1947.3+4447    & $11.6^{+1.7}_{-2.5}$       & $1.83^{+0.06}_{-0.17}$   & $0.29^{+0.45}_{-0.29}$   & $76^{+35}_{-32}$    & $237.1/217$   & 5              \\
    SWIFT J1952.4+0237    & $34.6^{+3.2}_{-5.3}$       & $1.71^{+0.09}_{-0.28}$   & $0.49^{+0.69}_{-0.47}$   & $100^{+49}_{-43}$   & $136.0/127$   & 5              \\
    SWIFT J2006.5+5619    & $26.0^{+2.8}_{-4.2}$       & $1.82^{+0.15}_{-0.21}$   & $0.42^{+0.63}_{-0.24}$   & $<93$               & $71.1/75$     & 5            \\
    SWIFT J2018.8+4041    & $19.2^{+0.8}_{-2.9}$       & $1.69^{+0.03}_{-0.18}$   & $0.73^{+0.32}_{-0.39}$   & $125^{+32}_{-36}$   & $194.4/214$   & 5              \\
    SWIFT J2052.0-5704    & $25.4 \pm 2.3$             & $1.59 \pm 0.15$          & $0.90^{+0.41}_{-0.32}$   & $198^{+28}_{-32}$   & $373.0/391$   & 5              \\
    SWIFT J2201.9-3152    & $12.6^{+0.3}_{-0.4}$       & $1.89^{+0.01}_{-0.03}$   & $0.62^{+0.10}_{-0.06}$   & $61^{+8}_{-9}$      & $1130.0/1102$ & 5              \\
    SWIFT J2359.3-6058    & $12.5^{+4.3}_{-4.2}$       & $1.41 \pm 0.27$          & $0.34^{+0.58}_{-0.34}$   & $189^{+58}_{-54}$   & $100.9/114$   & 5              \\
    
		\hline  
	\end{tabular}
	\tablefoot{The Swift BAT name of each source is listed in the first column. The best-fit absorption, photon index, and reflection strength are given in the next three columns, while the Fe line EW,  as estimated from the best-fit model, is listed in the fifth column. The penultimate column lists the $\chi^2$ statistic and the degrees of freedom of each fit. The stated upper limits correspond to 1-$\sigma$ level. All the values are retrieved from \pw \hspace*{-3pt}. The last column lists the BAT class of each source, as this is given in the 70-month \textit{Swift}/BAT catalogue.}
\end{table*}

\begin{figure}
  \centering
  \includegraphics[width=\linewidth,height=1.1\linewidth, trim={10 40 10 70}, clip]{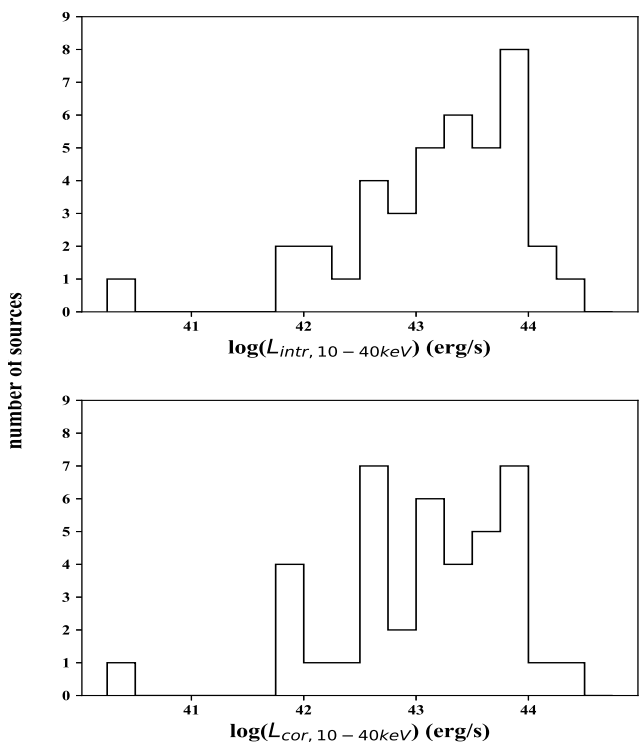}
  \caption[10]{Distribution of the intrinsic (top) and corona (bottom panel) luminosity in the 10-40 keV energy range. The luminosities have been estimated using the best-fit results of Sect. \ref{sect:pexrav_res} }
  \label{fig:lumin_hist}
\end{figure}

Using the best-fit results of the phenomelogical model we also estimated the radiated luminosity of each source, assuming an isotropic emission. The intrinsic AGN luminosity, that is the observed luminosity if no photo-electric absorption were present, and the corona luminosity, meaning the luminosity corresponding only to the power-law component, were estimated from 10 to 40 keV. Both luminosities were corrected for Compton scattering, which is not taken into account by the used \textit{zphabs} model, using the Thomson cross-section. The distribution of the calculated luminosities is plotted in Fig. \ref{fig:lumin_hist}.

\subsubsection{Using MYTorus}

In the following, the observed energy spectra were fitted by a more physical model. We chose to use the \textit{MYTorus} model \citep{2009MNRAS.397.1549M}, which has been frequently used in the past literature. This model simulates the effects of a homogeneous torus surrounding an isotropic central X-ray source. The source is assumed to radiate a power-law emission, while the torus has a half-opening angle of 60 degrees and covers half of the sky as observed from the central source. 

\textit{MYTorus} consists of three model components; the zeroth order continuum, called MYTZ, which simulates the modification of the intrinsic power law due to interaction of the photons with the torus matter, the scattered continuum emission, MYTS, which reproduces the spectrum of photons being scattered into our line of sight after interacting with the torus, and the fluorescent emission-line component, MYTL, which simulates the Fe K$\alpha$ and K$\beta$ lines produced by the torus at 6.4 and 7.06 keV, respectively. The \textit{MYTorus} model takes into account only photo-electric absorption and Compton scattering on free electrons and it assumes the element abundances of \cite{1989GeCoA..53..197A}.

The model parameters of \textit{MYTorus} are the photon index, $\Gamma^{\text{(MYT)}}$, and normalisation of the intrinsic power law, the equatorial column density of the torus, $N_\text{H, eq}^{\text{(MYT)}}$, and the source inclination, $\theta$. In addition, one may consider a cross-normalisation between MYTZ and MYTS or MYTL, which, if different from unity, may account for any anisotropy or variability of the central source. We chose to fit the spectra using the `coupled' version of \textit{MYTorus} \citep{2012MNRAS.423.3360Y}, which corresponds to the cross-normalisations being fixed at one, and thus, the emission interacting with the torus is assumed to be the same as the observed one. Finally, following the manual of \textit{MYTorus}, the fluorescent lines were smoothed using the convolution model \textit{gsmooth}, with a width of 0.05 keV.

The column density $N_\text{H, eq}^{\text{(MYT)}}$ corresponds to the column density of the torus when the source is observed edge-on. Its value largely determines the shape of the scattered spectrum. The column density along our line of sight can be estimated using the inclination angle and the following relation:

\begin{eqnarray}
\label{eq:nh_los}
  N_\text{H, LOS}^{\text{(MYT)}} = N_\text{H, eq}^{\text{(MYT)}} \left (1 - 4 \text{cos}^2\theta \right )^{1/2} .
\end{eqnarray}

\noindent It should be noted that the source inclination, as deduced by the fit, is valid only in the framework of the MYTorus model, and may not represent the real inclination of the source if the true geometry is very different from that used in MYTorus.

The best-fit values are given in Table \ref{tab:best_fit_mytorus}. The quality of the fit seems to be roughly similar to the one achieved with the phenomelogical model, even though each fit has two more degrees of freedom when \textit{MYTorus} is used. The average reduced $\chi^2$ statistic is now <$\chi^2$>$ = 0.98$, although it should be mentioned that the fit results in a null hypothesis probability $P_{null}<1\%$ for \textit{SWIFT J1325.4-4301} and \textit{SWIFT J1838.4-6524}. We did not perform a detailed statistical analysis to determine if either of the two models is highly preferred by the data, since most likely neither of the two corresponds to the real geometry of AGN in its full complexity. Instead, we were interested in whether the two models produce similar trends, which is shown to be true in the following section.

\begin{table*}
	\centering
	\caption{Best-fit parameters of the \textit{MYTorus} model.}
	\label{tab:best_fit_mytorus}
	\begin{tabular}{lllllr} 
		\hline
		{Source Name}         &  $N_\text{H, eq}^{\text{(MYT)}}$  &  $\Gamma$            &  $N_\text{H, LOS}^{\text{(MYT)}}$     & $\theta$                    &$\chi^2$/df    \\
		                      &  $(10^{24}$cm$^{-2})$             &                          &  $(10^{22}$cm$^{-2})$             &  (degrees)                  &                \\
		\hline
		SWIFT J0048.8+3155    &   $0.26_{-0.07}^{+0.04}$      & $1.60 \pm 0.01$            & $ 7.3^{+3.9}_{-2.6}$              & $61.35_{-0.59}^{+1.40}$      & 846.2/928   \\
    SWIFT J0152.8-0329    &   $6.66_{-2.39}^{+1.68}$      & $1.66 \pm 0.04$            & $ 54.3^{+52.1}_{-31.0}$           & $60.11_{-0.10}^{+0.20}$      & 96.7/99   \\
    SWIFT J0241.3-0816    &   $0.60_{-0.06}^{+0.12}$      & $1.73 \pm 0.01$            & $ 13.8^{+3.0}_{-2.5}$             & $60.88_{-0.27}^{+0.19}$      & 561.5/629   \\
    SWIFT J0505.8-2351    &   $1.99_{-0.46}^{+0.12}$      & $1.89 \pm 0.01$            & $ 25.3^{+42.0}_{-5.9}$            & $60.27_{-0.02}^{+0.90}$      & 540.1/551   \\
    SWIFT J0526.2-2118    &   $9.96_{-0.45}^{+0.04}$      & $1.83 \pm 0.03$            & $ 64.8^{+11.1}_{-3.4}$            & $60.07_{-0.01}^{+0.02}$      & 307.3/286   \\
    SWIFT J0623.9-6058    &   $2.00_{-0.52}^{+0.16}$      & $1.79 \pm 0.03$            & $ 31.7^{+10.1}_{-8.5}$            & $60.42_{-0.06}^{+0.26}$      & 151.9/180   \\
    SWIFT J0641.3+3257    &   $2.58_{-0.64}^{+0.87}$      & $1.80_{-0.02}^{+0.03}$     & $ 28.3^{+72.1}_{-12.9}$           & $60.20_{-0.15}^{+1.01}$      & 170.9/194   \\
    SWIFT J0742.3+8024    &   $0.12_{-0.02}^{+0.06}$      & $1.77 \pm 0.06$            & $ 12.0^{+5.5}_{-1.8}$             & $>68.22$                     & 117.5/117   \\
    SWIFT J0804.2+0507    &   $2.14_{-0.18}^{+0.08}$      & $1.66 \pm 0.02$            & $ 32.8^{+6.0}_{-4.1}$             & $60.39_{-0.07}^{+0.14}$      & 367.9/322   \\
    SWIFT J0804.6+1045    &   $0.26_{-0.15}^{+0.16}$      & $1.83 \pm 0.05$            & $ 10.2^{+51.0}_{-7.0}$            & $62.66_{-1.89}^{+27.34}$     & 176.5/169   \\
    SWIFT J0843.5+3551    &   $0.22_{-0.02}^{+0.30}$      & $1.48\pm 0.08$             & $ 22.1^{+30.0}_{-1.7}$            & $>65.10$                     & 64.7/69   \\
    SWIFT J0855.6+6425    &   $4.93_{-0.76}^{+0.79}$      & $1.72_{-0.06}^{+0.05}$     & $ 36.4^{+22.4}_{-11.7}$           & $60.09_{-0.05}^{+0.11}$      & 63.4/61   \\
    SWIFT J0902.0+6007    &   $8.92_{-1.09}^{+1.08}$      & $1.85 \pm 0.07$            & $ 62.0^{+38.5}_{-11.2}$           & $60.08_{-0.02}^{+0.10}$      & 58.9/47   \\
    SWIFT J0926.1+6931    &   $1.65_{-0.65}^{+0.41}$      & $1.82_{-0.02}^{+0.03}$     & $ 21.4^{+9.6}_{-9.3}$             & $60.28_{-0.10}^{+0.21}$      & 245.8/254   \\
    SWIFT J1044.1+7024    &   $6.92_{-2.06}^{+0.73}$      & $1.78 \pm 0.05$            & $ 51.0^{+40.5}_{-18.2}$           & $60.09_{-0.04}^{+0.14}$      & 90.5/88   \\
    SWIFT J1046.8+2556    &   $0.88_{-0.27}^{+0.40}$      & $1.58 \pm 0.06$            & $ 14.5^{+13.7}_{-6.3}$            & $60.45_{-0.28}^{+0.74}$      & 70.5/69   \\
    SWIFT J1049.4+2258    &   $0.33_{-0.04}^{+0.32}$      & $1.58 \pm 0.05$            & $ 30.8^{+30.1}_{-8.0}$            & $79.69_{-16.13}^{+10.31}$    & 184.4/195   \\
    SWIFT J1105.7+5854A   &   $2.96_{-0.63}^{+1.78}$      & $1.74 \pm 0.07$            & $ 33.3^{+49.7}_{-14.4}$           & $60.21_{-0.16}^{+0.57}$      & 41.8/48   \\
    SWIFT J1217.2-2611    &   $0.07_{-0.01}^{+0.09}$      & $1.55_{-0.03}^{+0.04}$     & $ 6.7^{+9.2}_{-0.7}$              & $>63.43$                     & 162.0/177   \\
    SWIFT J1219.4+4720    &   $0.34_{-0.16}^{+0.06}$      & $1.87_{-0.03}^{+0.02}$     & $ 9.1^{+13.9}_{-4.9}$             & $61.19_{-0.62}^{+3.65}$      & 567.5/569   \\
    SWIFT J1325.4-4301    &   $0.21_{-0.03}^{+0.01}$      & $1.79 \pm 0.01$            & $ 8.6^{+1.6}_{-1.2}$              & $62.88_{-0.26}^{+1.07}$      & 2941.0/2691   \\
    SWIFT J1341.5+6742    &   $2.53_{-0.57}^{+0.72}$      & $1.86_{-0.03}^{+0.02}$     & $ 27.8^{+15.2}_{-9.0}$            & $60.20_{-0.09}^{+0.19}$      & 164.9/163   \\
    SWIFT J1354.5+1326    &   $0.23_{-0.02}^{+0.10}$      & $1.52_{-0.08}^{+0.10}$     & $ 22.8^{+10.2}_{-1.7}$            & $>68.6$                      & 53.7/72   \\
    SWIFT J1457.8-4308    &   $3.00_{-0.67}^{+0.66}$      & $1.62 \pm 0.04$            & $ 30.4^{+13.1}_{-9.1}$            & $60.17_{-0.07}^{+0.13}$      & 90.5/98   \\
    SWIFT J1515.0+4205    &   $5.11_{-0.38}^{+0.40}$      & $1.86 \pm 0.02$            & $ 33.2^{+10.7}_{-9.6}$            & $60.07 \pm 0.04$             & 281.6/250   \\
    SWIFT J1621.2+8104    &   $1.00 \pm 0.49$             & $1.86 \pm 0.06$            & $ 20.4^{+17.5}_{-10.8}$           & $60.69_{-0.28}^{+0.98}$      & 67.5/73   \\
    SWIFT J1717.1-6249    &   $3.98_{-0.07}^{+0.25}$      & $1.78 \pm 0.01$            & $ 40.3^{+3.5}_{-4.2}$             & $60.17_{-0.04}^{+0.02}$      & 1131.4/1102   \\
    SWIFT J1824.2+1845    &   $2.77_{-0.83}^{+1.29}$      & $1.64 \pm 0.04$            & $ 28.0^{+22.4}_{-13.0}$           & $60.17_{-0.12}^{+0.22}$      & 96.6/120   \\
    SWIFT J1824.3-5624    &   $1.99_{-0.13}^{+0.09}$      & $1.68_{-0.03}^{+0.02}$     & $ 32.7^{+5.7}_{-3.1}$             & $60.45_{-0.06}^{+0.15}$      & 247.1/222   \\
    SWIFT J1826.8+3254    &   $0.99_{-0.31}^{+0.06}$      & $1.91 \pm 0.02$            & $ 14.5^{+2.5}_{-5.0}$             & $60.36_{-0.10}^{+0.12}$      & 341.5/391   \\
    SWIFT J1830.8+0928    &   $0.48_{-0.31}^{+0.09}$      & $1.76_{-0.09}^{+0.10}$     & $ 14.8^{+67.5}_{-11.7}$           & $61.59_{-1.53}^{+14.78}$     & 69.2/75   \\
    SWIFT J1838.4-6524    &   $2.28_{-0.02}^{+0.03}$      & $1.84 \pm 0.01$            & $ 33.2^{+2.3}_{-2.3}$             & $60.35 \pm 0.05$             & 1362.2/1181   \\
    SWIFT J1913.3-5010    &   $1.97 \pm 0.40$             & $1.70 \pm 0.03$            & $ 31.0^{+14.7}_{-8.5}$            & $60.41_{-0.15}^{+0.35}$      & 155.6/191   \\
    SWIFT J1947.3+4447    &   $0.53_{-0.21}^{+0.23}$      & $1.77_{-0.03}^{+0.04}$     & $ 10.0^{+10.3}_{-5.0}$            & $60.59_{-0.38}^{+1.11}$      & 236.2/219   \\
    SWIFT J1952.4+0237    &   $1.31_{-0.17}^{+0.09}$      & $1.67_{-0.13}^{+0.04}$     & $ 28.8^{+54.1}_{-7.9}$            & $60.81_{-0.39}^{+3.07}$      & 134.8/129   \\
    SWIFT J2006.5+5619    &   $0.18_{-0.01}^{+0.41}$      & $1.66 \pm 0.09$            & $ 17.5^{+40.8}_{-1.6}$            & $88.79_{-27.30}^{+1.00}$     & 71.5/77   \\
    SWIFT J2018.8+4041    &   $2.84_{-0.85}^{+0.72}$      & $1.68 \pm 0.03$            & $ 32.7^{+18.6}_{-12.7}$           & $60.22_{-0.11}^{+0.22}$      & 194.9/216   \\
    SWIFT J2052.0-5704    &   $1.82_{-0.27}^{+0.14}$      & $1.75 \pm 0.02$            & $ 32.2^{+10.5}_{-5.5}$            & $60.52_{-0.09}^{+0.33}$      & 384.3/393   \\
    SWIFT J2201.9-3152    &   $2.72_{-0.06}^{+0.23}$      & $1.89 \pm 0.01$            & $ 25.9^{+2.9}_{-2.1}$             & $60.15 \pm 0.02$             & 1131.5/1104   \\
    SWIFT J2359.3-6058    &   $0.83_{-0.16}^{+0.22}$      & $1.71_{-0.05}^{+0.09}$     & $ 15.9^{+21.8}_{-5.4}$            & $60.61_{-0.35}^{+1.65}$      & 104.9/116   \\
    
		\hline  
	\end{tabular}
	\tablefoot{The equatorial column density, the photon index, the column density along our line of sight, and the inclination are listed in the second, third, fourth, and fifth column, respectively. The given lower limits correspond to 1-$\sigma$ level. The last column lists the best-fit $\chi^2$ statistic and the degrees of freedom.}
\end{table*}


\subsection{Reflection-$N_\text{H}$ correlation}
\label{sect:rvsnh}

In \cite{2019A&A...626A..40P} we found that the reflection parameter $R$ was correlated with the observed column density for a small sample of obscured sources with $N_\text{H} > 5 \times 10^{22}\text{ cm}^{-2}$. The reflection was found to increase mainly for objects with $N_\text{H} > 15 \times 10^{22}\text{ cm}^{-2}$. Here, we constrained our analysis only to MOB sources with $N_\text{H} > 10^{23}\text{ cm}^{-2}$ to better characterise any correlation.

\begin{figure}
  \centering
  \includegraphics[width=\linewidth,height=0.8\linewidth, trim={20 10 20 30}, clip]{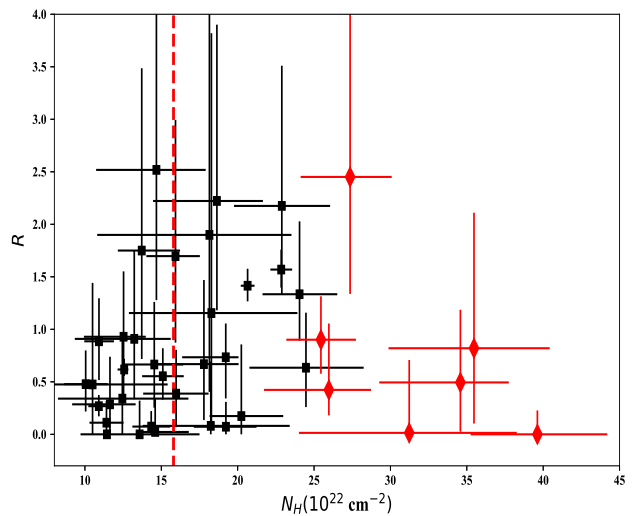}
  \caption[10]{Best-fit $R$ as a function of $N_\text{H}$. The sources denoted by a red diamond are excluded from the present analysis (see text for details). The vertical dashed red line denotes the boundary between the two groups of sources as defined in Sect. \ref{sect:rvsnh}.}
  \label{fig:rvsnh}
\end{figure}

The scatter plot of $R$ as a function of $N_\text{H}$ is shown in Fig. \ref{fig:rvsnh}. A visual examination of this plot does not allow a robust conclusion on the detection of any correlation. One may notice that the sources with the largest values of absorption tend to have low amount of reflection. However, this is simply a selection effect.  We excluded several reflection dominated sources\footnote{The term `reflection dominated' is used throughout this work to describe sources whose spectrum resembles roughly the shape of reflected emission.} from the considered sample when the softness ratio, defined as

\begin{eqnarray}
    SR &= \frac{CR_{3-5} - CR_{25-35}}{CR_{3-5} + CR_{25-35}},
\end{eqnarray}

\noindent was found to be smaller than 0.2 (see PW20, where this exclusion was conducted). The quantities $CR_{3-5}$ and $CR_{25-35}$ in the above equation denote the count rate in the energy range 3-5 and 25-35 keV, respectively. The excluded sources feature spectra mainly with large absorption and arbitrarily high reflection, which makes their fitting non trivial.

However, we estimated, now, that the softness ratio of even a moderately obscured source with reasonably high reflection might be below the threshold of 0.2. More precisely, we found that a source with $N_\text{H}= 35 \times 10^{22}\text{ cm}^{-2}$ and $R=1$ would result in a spectrum with $SR=0.07$, while for an observed emission with $N_\text{H}= 25 \times 10^{22}\text{ cm}^{-2}$ and $R=2$ the softness ratio would be $SR=0.17$. This highlights the possibility that sources with strong reflection in the high end of the considered absorption range have likely been excluded. Consequently, it is reasonable to assume that the lack of heavily obscured high-reflection sources is a side effect of our exclusion criterium. We therefore decided to limit the analysis to sources with $N_\text{H} < 25 \times 10^{22}\text{ cm}^{-2}$. The aforementioned effect is not significant for sources with $N_H$ below $25 \times 10^{22}\text{ cm}^{-2}$, as it is evident from Fig. \ref{fig:rvsnh}.

The reflection plotted in Fig. \ref{fig:rvsnh} seems to be increasing on average with the column density. We estimated \footnote{All the calculations of the statistical tests mentioned in this work were conducted with the \textit{python} library \textit{SciPy} (https://docs.scipy.org/doc/scipy/reference/index.html). } the Spearman rank correlation coefficient for the two variables to be $\rho = 0.37$ with a chance probability of $P_\text{null} = 3.2\%$. Motivated by this result, we proceeded in examining the apparent correlation with different statistical tests in order to conclude on its detection with high significance.

\begin{figure}
  \centering
  \includegraphics[width=\linewidth,height=0.7\linewidth, trim={0 0 0 30}, clip]{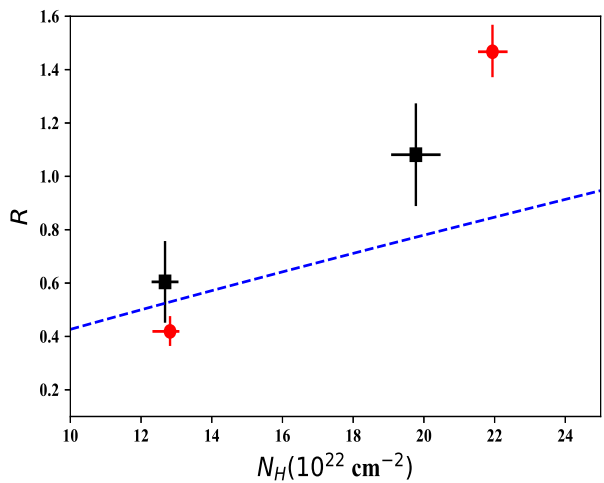}
  \caption[10]{Observed $R$-$N_\text{H}$ correlation. The black squares correspond to the average values of the individual fits of the two considered groups, while the red circles denote the best-fit values of the corresponding composite spectra. The blue dashed line indicates the detected correlation of \cite{2020ApJ...901..161K}.}
  \label{fig:rvsnh_stack}
\end{figure}

First, we divided the sources into two logarithmically equally spaced groups based on the best-fit $N_\text{H}$. The first group (MOB1) includes 18 sources with $10 \times 10^{22} \text{cm}^{-2} < N_\text{H} < 15.8 \times 10^{22} \text{cm}^{-2}$, and the second group (MOB2) comprises 15 sources with $15.8 \times 10^{22} \text{cm}^{-2} < N_\text{H} < 25 \times 10^{22} \text{cm}^{-2}$. As shown in Fig. \ref{fig:rvsnh_stack}, the average value of $R$ in MOB2 is much larger than the average reflection of MOB1. Using Welch's t-test the two groups were found to have the same mean $R$ with a null hypothesis probability of $P_\text{null} = 6.3\%$.

Moreover, we calculated the composite spectrum of each group by stacking the corresponding spectra. Our aim was to verify that the analysis of stacked spectra does not result in a significantly different trend. Before stacking, each individual spectrum was renormalised based on the average value of its ARF file in order to account for differences in the source extraction detector regions. The spectrum of \textit{SWIFT J1325.4-4301} was not considered in the composite spectrum of MOB1, because its signal-to-noise ratio was significantly (more than an order of magnitude) higher than that of any other spectrum. We then fitted the deduced stacked spectra using the phenomelogical model presented in Sect. \ref{sect:pexrav_res}. The best-fit $R$ versus $N_\text{H}$ are plotted in Fig. \ref{fig:rvsnh_stack}. Clearly, the reflection is increasing with absorption. The similarity of this trend to the one observed for the individual fits provides further confidence for the reality of the observed $R$-$N_\text{H}$ correlation. However, the significance of the reflection increase observed for the stacked spectra was not evaluated statistically in order to avoid any bias due to the stacking process.

Before proceeding, it should be mentioned that the two groups were examined for differences in the photon index, the intrinsic and corona luminosity, and the BAT luminosity from 14 to 195 keV using the Anderson-Darling test \citep{10.2307/2236446} \footnote{The Anderson-Darling test is a statistical test examining if two groups are drawn from the same parent distribution. It is similar to the frequently used Kolmogorov-Smirnov test, but it has been found to be more powerful \citep{2011razali}. An interesting discussion on this subject, focusing specifically in astronomy, is given in https://asaip.psu.edu/articles/beware-the-kolmogorov-smirnov-test/ .}. This was necessary to confirm that the $R$-$N_\text{H}$ correlation is not an artefact of an underlying correlation or a selection effect. No difference was found for the examined variables between the two groups with the corresponding null hypothesis probability $P_\text{null} > 10\%$ for every conducted test.

In conclusion, different statistical tests indicated that the reflection, as parametrised by $R$, is on average increasing with absorption. All the tests resulted in a chance probability of $P_\text{null} < 7\%$. However, none of the considered tests was significant at a $P_\text{null}<1\%$ level. 

In the following, we studied how the Fe K$\alpha$ emission line evolves with the absorption column density. Since the Fe line is also produced by the interaction of the intrinsic X-ray continuum with the surrounding matter, its study might provide further constraints about this matter in the various sources.

\begin{figure}
  \centering
  \includegraphics[width=\linewidth,height=0.7\linewidth, trim={0 0 0 30}, clip]{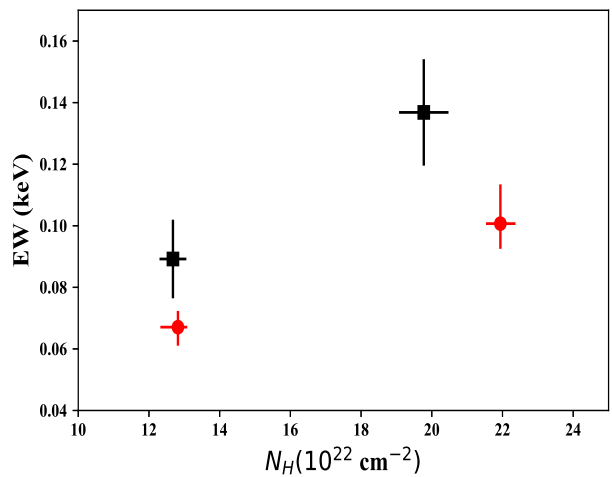}
  \caption[10]{Observed Fe EW-$N_\text{H}$ correlation in the two absorption groups. The black squares correspond to the average values of the individual fits of the two groups, while the red circles denote the best-fit values of the corresponding composite spectra.}
  \label{fig:eqwvsnh_stack}
\end{figure}

Figure \ref{fig:eqwvsnh_stack} plots the Fe K$\alpha$ EW as a function of $N_\text{H}$ for the two groups, MOB1 and MOB2, defined previously. The EW is increasing on average with $N_\text{H}$, similarly to $R$. A Welch's t-test gave a probability of $P_\text{null}=3.5\%$ for the two groups to have the same average EW.

The increase in both Fe EW and $R$ with $N_\text{H}$ suggests that the total amount of reflected emission is higher in MOB2 sources, which might indicate an increase in the total quantity of matter around the central source. More importantly, the Fe EW and $R$ are independent in the used model, since they result from different model components \footnote{The independence of $R$ and EW was also verified when using the \textit{steppar} command in \textit{XSPEC} to study how EW varies for different values of $R$. No specific trend was found.}. Therefore, the two correlations, $R$ versus $N_\text{H}$ and EW versus  $N_\text{H}$, are independent.

We calculated the chance probability that both parameters, $R$ and EW, increase in the MOB2 group. To do so, we assumed that the observed $R$ and EW distributions of all the considered sources are equal to the parent distributions from which the values for both MOB1 and MOB2 sources are drawn. Then we randomly picked 18 pairs of values, which are attributed to the first simulated group of values, and 15 pairs of values for the second group. This process is repeated one million times. After every run we computed the difference between the average reflections and the average EWs of the two simulated groups. We estimated the requested chance probability by dividing the number of runs for which the difference between the average values of reflection of the simulated groups is larger than the observed difference in R between MOB1 and MOB2, and, simultaneously the difference between the average values of EW of the simulated groups is larger than the observed EW difference between MOB1 and MOB2, by the number of total runs. In this way, the chance probability was found to be $P_\text{null} = 0.1\%$.

\begin{figure}
  \centering
  \includegraphics[width=\linewidth,height=0.7\linewidth, trim={0 0 0 30}, clip]{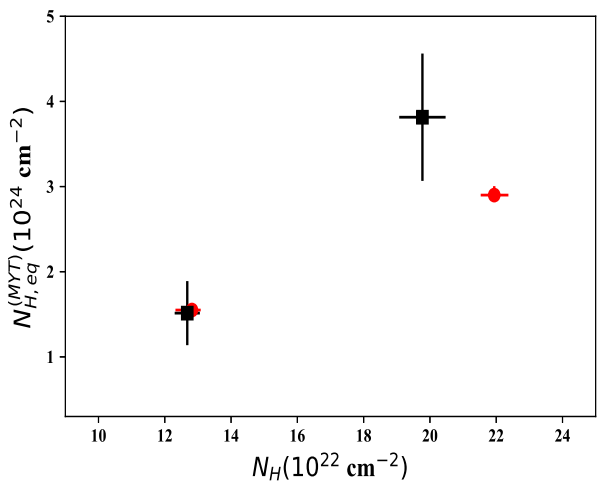}
  \caption[10]{Observed $N_\text{H, eq}^{\text{(MYT)}}$ versus $N_\text{H}$ in the two absorption groups. The black squares correspond to the average values of the individual fits for the two groups, while the red circles denote the best-fit values of the corresponding composite spectra.}
  \label{fig:nhmytvsnh_stack}
\end{figure}

Finally, we explored how the equatorial column density, derived from the \textit{MYTorus} model, changes from MOB1 to MOB2. The scatter plot of $N_\text{H, eq}^{\text{(MYT)}}$ versus $N_\text{H}$ for the two groups is plotted in Fig. \ref{fig:nhmytvsnh_stack}. The equatorial column density increases on average with $N_\text{H}$ as well. Using the individual best-fit values, the two parameters were found to be correlated with a Spearman rank correlation coefficient of $\rho = 0.43$ and a chance probability of $P_\text{null} = 1.3\%$.

This result further supports that the reflection is stronger in MOB2 sources. An increased $N_\text{H, eq}^{\text{(MYT)}}$ indicates an increased curvature of the observed energy spectrum in energies above 10 keV, in excess to the curvature expected due to an increase in the line of sight $N_\text{H}$. The increased curvature is physically attributed to Compton reflection. This curvature is reproduced as an increase in the reflection strength when the phenomelogical \textit{pexrav} model is used and as an increase in equatorial column density within the \textit{MYTorus} model.

In terms of spectral shape, the model fitting results suggest that sources with larger curvature at low energies exhibit larger curvature at energies above 10 keV, as well. The higher curvature at low energies is naturally ascribed to an increase in absorption, while the greater curvature at high energies is associated with an increase in the reflected emission. To verify this trend in a model-independent way, we estimated the following softness ratios:

\begin{eqnarray}
\label{eq:sr1}
    SR_1 &= \frac{CR_{3-5} - CR_{8-12}}{CR_{3-5} + CR_{8-12}}, \\
\label{eq:sr2}
    SR_2 &= \frac{CR_{8-12} - CR_{15-25}}{CR_{8-12} + CR_{15-25}},
\end{eqnarray} 

\noindent where $CR_{8-12}$ and $CR_{15-25}$ are the count rates from 8 to 12 keV and from 15 to 25 keV, respectively. Defined in such a way, $SR_1$ is an estimation of the spectral shape in low energies, while $SR_2$ quantifies the spectral shape at higher energies. Keeping in mind that the softness ratios depend also on the photon index among other factors, one may visualise differences in $SR_1$ as a proxy for differences in $N_\text{H}$ and differences in $SR_2$ as a proxy for differences in $R$. $SR_1$ is expected to be decreasing when $N_H$ increases, while $SR_2$ would feature smaller values for larger $R$.

\begin{figure}
  \centering
  \includegraphics[width=\linewidth,height=0.7\linewidth, trim={0 0 0 30}, clip]{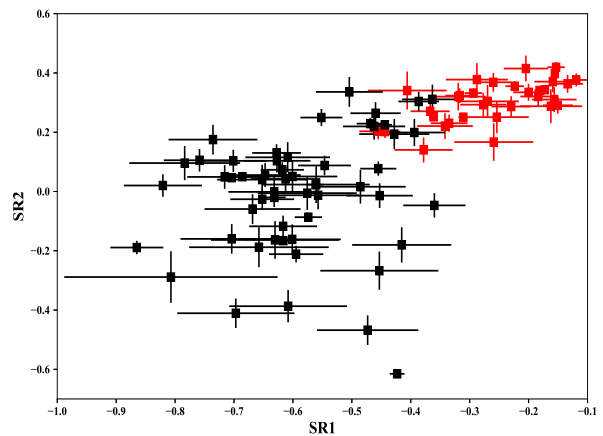}
  \caption[10]{Scatter plot of softness ratios SR1 and SR2, as defined in Sect. \ref{sect:rvsnh}. The red squares denote the MOB1 and MOB2 sources, while the black squares correspond to the objects excluded as reflection dominated (see Sect. \ref{sect:refl_dom}).}
  \label{fig:softness_ratios}
\end{figure}

Figure \ref{fig:softness_ratios} plots $SR_2$ as a function of $SR_1$. The two ratios are positively correlated, indicating that the spectral shapes at high and low energies are indeed correlated. Following the discussion above, Fig. \ref{fig:softness_ratios} provides extra evidence that $R$ scales with $N_H$.

In conclusion, the spectral fitting results of both considered models and the analysis of the spectral shape as quantified by the above softness ratios point towards the same trend. The physical interpretation of this trend is discussed in Sect. \ref{sect:discuss}.


\section{Discussion}

\label{sect:discuss}

\subsection{Reflection-$N_\text{H}$ correlation}

The reflection strength of the \textit{pexrav} model was found to be positively correlated to the observed column density. The validity of this correlation was also verified when using a more physical model for the spectral fitting, namely \textit{MYTorus}. Such a correlation was recently observed by \cite{2020ApJ...901..161K} as well. However, the trend found here indicates a steeper correlation than the one found by that work, although the results are roughly consistent (Fig. \ref{fig:rvsnh_stack}). The flatter correlation found by \cite{2020ApJ...901..161K} might be due to the fact that they analysed sources within a larger range of $N_\text{H}$, and thus the effect observed in our work might be smoothed out by effects in other absorption states or by selection effects.

The detected correlation is confirmed by the fact that the Fe K$\alpha$ EW was also increasing on average with $N_\text{H}$. The Fe K$\alpha$ emission line is, similarly to the Compton hump, produced by the interaction of the intrinsic X-ray continuum with the matter surrounding the central black hole. It should be noted, though, that the Fe EW and $R$ are not expected a priori to follow a similar trend. Apart from potential differences of the iron abundance in the studied AGN, $R$ and Fe EW may differ because the iron line may be produced from Compton thin material, as well, while the Compton hump requires a Compton thick (CT) source. For instance, \cite{2017ApJ...842L...4F} have found that substantial amount of the observed Fe line is due to gas of the host galaxy, several kpc away from the nucleus.

Nevertheless, the detection of an increase in the average of both $R$ and Fe EW as $N_\text{H}$ increases with a chance probability of $P_\text{null}=0.1\%$ indicates that the total reflected emission is strongly associated with the line of sight $N_\text{H}$. This result cannot be easily explained by the simple unification model. For example, assuming a homogeneous torus with similar matter density in every AGN, $N_\text{H}$ will increase as the source inclination increases. In this case, the reflection strength is expected to be reduced as smaller area of the reflected region is observed directly. Hence, the two quantities, $R$ and $N_\text{H}$, would be anti-correlated, contrarily to what is observed. A homogeneous torus might explain the observed trend if its matter density varies significantly between the sources, although this would require that sources with similar line of sight absorption differ significantly in the density of the surrounding gas (and potentially in their inclination) by a factor of a few (Fig. \ref{fig:nhmytvsnh_stack}).

The observed trend indicates that the covering factor of the reflecting matter is increasing with absorption. Such an increase can be naturally explained when a clumpy torus is assumed to surround the central source, as has also been supported by independent studies \citep{2014MNRAS.439.1403M}. In this case, the gas responsible for both the absorption and the reflection of the intrinsic emission resides in clouds rotating around the black hole. An $R$-$N_\text{H}$ correlation could be driven by an increase in the filling factor of the clouds.

\subsubsection{Reproducing the $R-N_\text{H}$ correlation}

We examined this assumption using the \textit{RefleX} software \citep{2017A&A...607A..31P}, a ray-tracing code that simulates the propagation of X-rays through matter with arbitrary geometry. We assumed a central point source emitting isotropically following a power law with a photon index of 1.8 and a cutoff of 200 keV, surrounded by a clumpy torus. The radial boundaries of the torus were fixed so that the outermost boundary is 100 times larger than the innermost boundary. The spatial distribution of clouds was assumed to have an azimuthal symmetry. Following \cite{2008ApJ...685..160N}, the clouds altitude distribution was given by:

\begin{eqnarray}
\label{eq:clouds_distro}
  N = N_0 \cdot \exp \left [ - \left (\frac{\omega}{\sigma}  \right )^{2} \right ] ,
\end{eqnarray}

\noindent where $\omega$ is the elevation angle of the cloud as measured from the edge-on direction, $N$ is the average number of clouds towards the $\omega$ direction and $N_0$ is the average number of clouds along the equatorial. The angular dispersion of the clouds, $\sigma$, was chosen to be 20 degrees as measured from the equatorial direction. Following \cite{2019A&A...629A..16B}, we consider the angular size of each cloud as observed by the central source, which we assumed to follow a Gaussian distribution with a mean of one degree and a standard deviation of half a degree. The matter in the clouds was selected to be neutral with Solar metallicity. Finally, the absorption column density through the centre of each cloud was fixed at $N_\text{H, cloud} = 10^{23} \text{cm}^{-2}$, which is similar to the values found by \cite{2014MNRAS.439.1403M}.

It should be mentioned that the problem of a clumpy torus has a high degree of degeneracy, meaning that different values of the above parameters might provide similar results. A detailed consideration of the whole parameter space is outside the scope of this work and we only aim at studying the effect of the clouds filling factor, with the remaining parameters fixed at reasonable values. We expect our results to be representative of a large fraction of the realistic space of parameters and distributions. \cite{2019A&A...629A..16B} discuss the effects of the different parameters, while \cite{2020ApJ...897....2T} have found that the spectra of obscured AGN show a large variety of clumpy torus's properties.

The only parameter left to determine the problem uniquely is the filling factor, or equivalently the total number of clouds. We estimated the expected observed emission for different values of the filling factor. The typical number of clouds used was of the order of $10^5$. For each value, we were able to measure both the reflected and the absorbed continuum emissions, which were summed over all azimuthal angles. The simulated results are not directly comparable to the best-fit $R$ values. In order to compare the outcome of the \textit{RefleX} simulations to our results we estimated the reflection fraction, defined as the ratio of reflected emission over continuum emission in the 10-40 keV energy range.

\begin{figure}
  \centering
  \includegraphics[width=\linewidth,height=0.7\linewidth, trim={0 0 0 30}, clip]{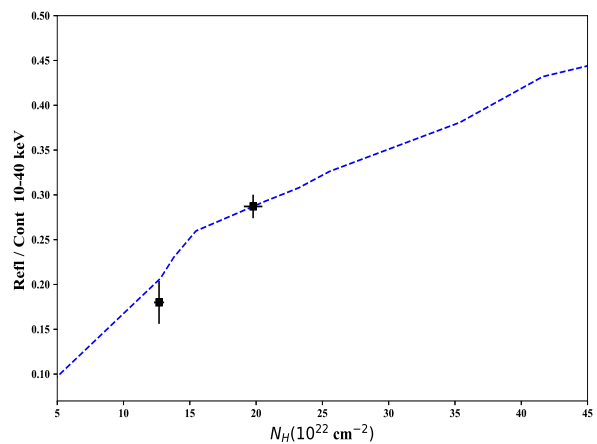}
  \caption[10]{Fraction of reflected over continuum emission from 10 to 40 keV as a function of absorption. The $y$ values of the black points were estimated as the average reflection fraction of the individual \textit{MYTorus} best-fit results. The blue dashed line denotes the predicted fraction in the case of a clumpy torus, as estimated by \textit{RefleX}. The small break at  $N_\text{H} \simeq 14 \times 10^{22}\text{cm}^{-2}$ is probably due to computational noise. }
  \label{fig:reflvsnh_reflex}
\end{figure}

Figure \ref{fig:reflvsnh_reflex} plots the predicted fraction for an inclination of 53 degrees as a function of $N_\text{H}$, which corresponds to the average $N_\text{H}$ value for the specific inclination over all the azimuthal directions. The same fraction was estimated for the data using the \textit{MYTorus} results. For each source, the reflection fraction was calculated using the best-fit results. The black points in Fig. \ref{fig:reflvsnh_reflex} indicate the average value of the reflection fraction in the MOB1 and MOB2 class versus the corresponding column density estimated in Sect. \ref{sect:rvsnh}.

The \textit{RefleX} results indicate that a positive $R$-$N_\text{H}$ correlation is expected under the assumption of a clumpy torus with a varying filling factor. The predicted relation matches well the observed one. The MOB1 and MOB2 values in Fig. \ref{fig:reflvsnh_reflex} correspond to a filling factor of nearly 0.16 and 0.28, respectively. The agreement between the predicted and observed trend suggests that a clumpy torus can explain the $R$-$N_\text{H}$ correlation detected in obscured AGN. The torus seems to be the main reflector and absorber in these sources.

The \textit{RefleX} predicted curve in Fig. \ref{fig:reflvsnh_reflex} also suggests that the $R$-$N_\text{H}$ correlation should extend to less absorbed objects, that is sources with $N_\text{H}< 10^{23} \text{cm}^{-2}$ would feature smaller values of $R$ as well. However, \pw have shown that less obscured sources, with $N_\text{H}< 10^{23} \text{cm}^{-2}$, feature similar levels of reflection to those of the more obscured objects, with $N_\text{H}>10^{23} \text{cm}^{-2}$ (see Fig. 5 in that paper). This implies that the reflection in less absorbed sources is not driven by the torus's filling factor, which might be explained by differences in the origin of the reflected emission or if differences in other physical parameters, such as the source inclination, become important.

Moreover, Fig. \ref{fig:reflvsnh_reflex} indicates that more absorbed sources, with $N_\text{H}>25 \times 10^{22} \text{cm}^{-2}$, are predicted to feature larger values of reflection. Such a trend seems to be supported by the data and is further discussed in the next section.


\subsection{Connection to reflection dominated AGN}
\label{sect:refl_dom}

As already mentioned, several sources have been excluded from the present analysis as reflection dominated. In Sect. \ref{sect:rvsnh} we excluded 7 sources, while PW20 excluded 50 objects of the initial sample. These sources correspond mostly to heavily absorbed, potentially CT, AGN. The study of these sources is not straightforward since it is challenging to distinguish the primary continuum from the reflected emission in their spectrum. In fact, initial fitting of these sources indicated that their spectrum is not well reproduced by any of the considered models (i.e. \textit{MYTorus} or \textit{pexrav}). 

Nevertheless, it is interesting to consider how these heavily obscured sources are connected to MOB objects. It is tempting to assume that the $R$-$N_\text{H}$ correlation observed in MOB objects can be extrapolated to sources of larger absorption. In this case, AGN with large $N_\text{H}$ (above $30 \times 10^{22} \text{cm}^{-2}$ or so) will feature a strong reflected emission in their spectrum. Indeed, \cite{2019A&A...626A..40P} found that the spectra of heavily absorbed AGN are dominated by the reflection component. In order to examine if this is the case for the current sample as well, we estimated the softness ratios, $SR_1$ and $SR_2$, as defined in eq. \ref{eq:sr1} and \ref{eq:sr2}, for the excluded sources. The estimated values are plotted in Fig. \ref{fig:softness_ratios}.

As stated previously, due to their definition, $SR_1$ and $SR_2$ can be viewed, at first approximation, as a proxy for $N_\text{H}$ and $R$, respectively. As a result, Fig. \ref{fig:softness_ratios} indicates that the reflection dominated sources do, indeed, feature larger values of $N_\text{H}$ and $R$ than those of MOB sources. However, the detection of any underlying correlation for the reflection dominated sources is not possible due to the large scatter of values.

Initial efforts to model the spectrum of reflection dominated sources suggested that the increase in $R$ as a function of $N_\text{H}$ in these objects is smaller than the one expected when the MOB correlation is extrapolated to higher absorption states. This potentially suggests a different geometry for the reflecting surface between the two groups or a saturation of reflection due to geometric effects. A similar result was obtained by \cite{2019A&A...629A..16B}, who found that a high-density reflector with large covering factor near the central source is needed to explain the X-ray spectrum of heavily absorbed nearby AGN.

It should be noted that further analysis is required before firm conclusions can be derived for the reflection dominated sources. Increased quality of spectra would be crucial for such an analysis, potentially combining deeper \textit{NuSTAR} observations with observations in softer X-rays. Additionally, a detailed consideration of the full parameter space in the case of a clumpy torus might reveal the regime within which these sources reside. Further information about these sources might also be obtained with the forthcoming polarimetry missions, \textit{IXPE} and \textit{eXTP} \citep{2016ResPh...6.1179W, 2016SPIE.9905E..1QZ}.

\subsection{Implications for the CXB}

The cosmic X-ray background \citep[CXB,][]{1962PhRvL...9..439G} is believed to be the integrated emission of AGN. Its spectral shape peaks at around 30 keV pointing to a strong contribution by highly absorbed AGN. This led to several studies \citep[e.g.][]{2007A&A...463...79G} trying to reproduce its exact spectrum following a population synthesis model, since such an approach could, in principle, derive limits on the amount of CT AGN in the Universe, which are not easily identified via direct observations.

However, \cite{2009ApJ...696..110T} have shown that such an approach includes several degeneracies, which could bias the final results. More precisely, \cite{2012A&A...546A..98A} showed that the reflection strength in AGN spectra is highly degenerated with the fraction of CT sources. Therefore, an accurate estimation of the CT fraction requires a good knowledge of the spectral characteristics of AGN. Otherwise, surveys at energies larger than 10 keV, which are less biased with respect to the absorption effects in detecting AGN, may provide more reliable constraints on the CT fraction. Recently, \cite{2016A&A...590A..49E} analysed the average BAT spectra of a large sample of nearby AGN. After constraining the spectral properties at different absorption levels, these authors found that the strong reflection observed in mildly obscured sources results in the number of CT sources to be equal to the value directly observed in X-ray surveys, of around 20\% of the total AGN population \citep{2012MNRAS.423..702B, 2015ApJ...815L..13R}. 

Although the $R$-$N_\text{H}$ correlation is expected to have important implications in estimating the CT fraction by reproducing the observed CXB, a detailed synthesis analysis using our best-fit results is outside the scope of this work. Nevertheless, we checked if our results are consistent with those obtained by \cite{2016A&A...590A..49E}. A direct comparison of the best-fit parameters is not possible due to the differences in the used model and in element abundances, as well as differences in the used sample. We, instead, cross-matched our PW20 sample with the one of \cite{2016A&A...590A..49E} and estimated the average \textit{NuSTAR} spectra for the three obscuration classes defined in their paper: the lightly obscured LOB1 sources with $10^{21} < N_\text{H} < 10^{22} \text{cm}^{-2}$, the lightly obscured LOB2 sources with $10^{22} < N_\text{H} < 10^{23} \text{cm}^{-2}$, and the mildly obscured sources considered in \cite{2016A&A...590A..49E}, MOB$_{EW16}$ with $10^{23} < N_\text{H} < 10^{24} \text{cm}^{-2}$. Out of the 90 sources in the original 2016 paper, the \textit{NuSTAR} spectra of 59 sources were studied in PW20. We, then, compared the stacked \textit{NuSTAR} spectra to the stacked BAT spectra of \cite{2016A&A...590A..49E}, since the main driver in reproducing the CXB is the spectral curvature above 10 keV in the different obscuration classes, regardless of the exact best-fit values of the model parameters. 

\begin{figure}
  \centering
  \includegraphics[width=\linewidth,height=0.7\linewidth, trim={0 0 0 30}, clip]{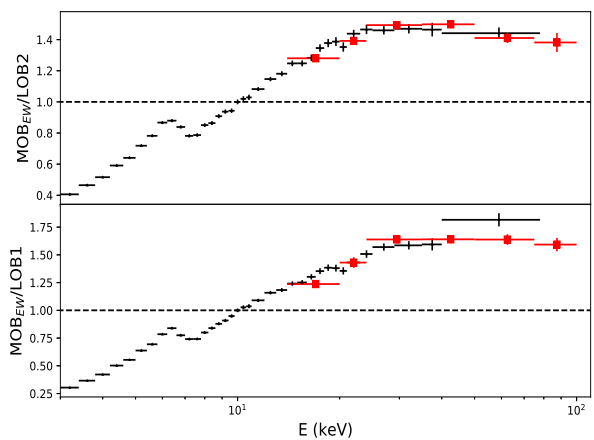}
  \caption[10]{Spectral ratio between the average spectra of MOB$_{EW16}$ over LOB1 (lower) and LOB2 (upper panel) classes, as these were defined in \cite{2016A&A...590A..49E}. The black points correspond to the \textit{NuSTAR} ratio and the red squares denote the BAT results. The horizontal dashed line indicates the $y=1$ line.}
  \label{fig:mob_over_lob12}
\end{figure}

Figure \ref{fig:mob_over_lob12} plots the ratio between the MOB$_{EW16}$ average spectrum and the LOB1 and LOB2 average spectra. It should be mentioned that before dividing, the \textit{NuSTAR} spectra were normalised to have the same value at 10 keV, while the BAT ratios are scaled to be directly comparable to the \textit{NuSTAR} ratio. As it is evident, the spectral curvature found by \textit{NuSTAR} is in agreement to the one derived when the BAT data are considered. Consequently, the results of \cite{2016A&A...590A..49E} seem to be supported by the \textit{NuSTAR} data, meaning that a large fraction of CT sources is not needed to explain the observed CXB.

A similar conclusion was reached by \cite{2016A&A...594A..73A}, who followed a Bayesian approach to determine the absorption column density of heavily obscured AGN in the 70-month \textit{Swift}/BAT catalogue. These authors deduced that CT sources account for about 10-20\% of the AGN population, while a significantly larger amount of CT AGN could only be reached if evolution of their fraction with redshift is assumed. Finally, \cite{2018ApJ...854...49M} studied the \textit{NuSTAR} spectra of CT candidates selected from the 100-month \textit{Swift}/BAT catalogue and found that the inclusion of high-quality \textit{NuSTAR} spectra decreases the amount of CT AGN in the local universe to values as low as 4\%.


\section{Conclusions}
\label{sect:conclud}

We studied the \textit{NuSTAR} spectra of nearby heavily absorbed AGN. The unique capabilities of \textit{NuSTAR} allowed us to investigate the various emission components and the absorption effects. Our main findings can be summarised as follows:

\begin{itemize}
  \item In the case of MOB ($10^{23} \text{cm}^{-2} < N_\text{H} < 2.5 \times 10^{23} \text{cm}^{-2}$) sources, the amount of reflected emission, as quantified by the reflection strength and the Fe EW, is positively correlated to the observed column density, $N_\text{H}$, indicating that the torus is the main reflector in these objects. This result was evident regardless of the used model, \textit{pexrav} or \textit{MYTorus}. The reflection seems to increase even further in the case of more heavily absorbed objects.
  \item The observed correlation is well reproduced as a result of increasing the torus filling factor, when the effects of a clumpy torus are modelled by the ray-tracing software \textit{RefleX}.
  \item The detection of strong reflection in absorbed sources suggests that a large fraction of CT AGN are not needed to explain the CXB, consistent with previous studies.
\end{itemize}

Our results are consistent with the existence of a clumpy torus around the central engine of AGN. Populating the observed correlation with more sources in the desired range of $N_\text{H}$ will allow the torus's characteristics to be constrained by testing different configurations. Similar explorations could be conducted if the studied sources are observed with larger exposure reducing thus the statistical uncertainties. Such studies are invaluable in infering the outer geometry of AGN, where the torus lies, and hence, in understanding how the accretion disc is supplied with matter during the active phase of AGN.

\bibliographystyle{aa} 
\bibliography{arxiv_draft} 

\end{document}